\documentstyle[epsf,prl,twocolumn,aps]{revtex}
\begin{document}
\title{A two-channel Kondo impurity in the spin-1/2 chain: 
Consequences for Knight shift experiments}
\author{Sebastian Eggert and Stefan Rommer}
\address{Institute of Theoretical Physics,
Chalmers and G\"oteborg University,
S-412 96 G\"oteborg, Sweden}
\maketitle
\begin{abstract}
A magnetic impurity in the spin-1/2 chain is a simple realization of
the two-channel Kondo problem since the field
theoretical descriptions in the spin-sector are identical.  
The correlation functions near the impurity can be calculated.
Using  a modified version
of the numerical transfer matrix DMRG, we are now able 
to accurately determine local properties close to the impurity 
in the thermodynamic limit.  The local 
susceptibilities (Knight-shifts) show an interesting behavior in a 
large range around the impurities. We are able to make
quantitative experimental predictions which would allow to observe two-channel
Kondo physics for the first time directly by doping of spin-1/2 chain
compounds.
\end{abstract}
We are considering a defect in the antiferromagnetic spin-1/2
chain, consisting of two altered bonds in the chain
\begin{equation}
H = J \sum_{i=1}^{L-1} {\bf S}_i \cdot {\bf S}_{i+1} \  +  \
J'({\bf S}_L  + {\bf S}_1 ) \cdot {\bf S}_0 . \label{ham}
\end{equation}
Such an impurity could be created in quasi-one dimensional 
spin compounds by doping with other spin-1/2 ions or substituting 
suitable non-magnetic neighboring atoms.  This paper represents
a short summary of the research for presentation purposes.  More
detailed calculations have been published elsewhere\cite{prl}.

From field theory calculations
it is known that the model in Eq.~(\ref{ham}) is equivalent to 
the two-channel Kondo problem in terms of the renormalization 
behavior\cite{eggert1,clark,affleck}.   
For a weak coupling $J'$ the impurity spin ${\bf S}_0$
is almost free, but the effective coupling
grows logarithmically as the temperature is lowered. 
The system then renormalizes to a stable intermediate coupling fixed 
point $J' = J$ which corresponds to the ``healed'' chain.
Our calculations now give quantitative results which can be compared
to experiments, so that the two-channel Kondo physics can be
observed directly. For our calculations we have used the numerical 
transfer matrix renormalization group\cite{TMDMRG}, 
which we have modified to calculate impurity properties directly
in the thermodynamic limit.

\begin{figure}
\begin{center}
\mbox{\epsfxsize=3.35in \epsfbox{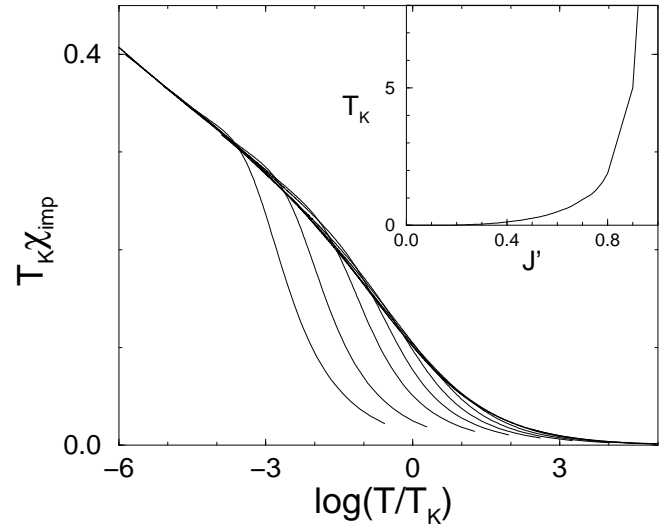}}
\end{center}
\caption{The scaled impurity susceptibility $\chi_{\rm imp}$,
which shows logarithmic behavior and the scaling of
Eq.~(\protect{\ref{scaling}}) for an
appropriate choice of $T_K$ as a function of $J'= 0.1J, \ ...,\ 0.95J$ (inset).
\label{chiimp}}
\end{figure}

In particular, we consider the local susceptibilities over a 
large range around the impurity as a function of site index $x$
\begin{equation}
\chi(x) \ = \ \left.\frac{d \langle S^z(x)\rangle_B}{dB}
\right|_{B=0},
\end{equation}
where $B$ is a uniform magnetic field on the entire chain.
The susceptibility of the impurity spin $\chi(0)$ is a good
approximation to the true impurity susceptibility
\begin{equation}
\chi_{\rm imp} \approx \chi(0)  - J' \chi,
\end{equation}
where we have subtracted the pure susceptibility per site $\chi$ to only
capture the impurity effects.   We
 find that the impurity susceptibility $\chi_{\rm imp}$
is logarithmic divergent with temperature as expected from the
two-channel Kondo effect.  Moreover, we can explicitly show that 
the impurity susceptibility indeed follows the  scaling form 
\begin{equation}
\chi_{\rm imp}(T) = f(T/T_K)/T_K,  \label{scaling}
\end{equation}
as shown in Fig. \ref{chiimp} up to deviations at higher
temperatures which are due to non-universal behavior above the cut-off.  
Here $T_K$ is the crossover temperature, which is exponentially small 
near the unstable fixed point $J' \ll J$ and diverges at the stable
fixed point with $(J-J')^{-2}$.

More interesting is the spatial dependence of $\chi(x)$ as a function 
of site index.  It is known that open boundary conditions induce a large
staggered part which {\it increases} with the distance from the boundary
$\chi(x) \propto  (-1)^{x} {x \sqrt{T}}/{\sqrt{\sinh 4 x T}},$
where $T$ is measured in units of $J$.  
This surprising effect has been confirmed by NMR experiments\cite{NMR}.

\begin{figure}
\begin{center}
\mbox{\epsfxsize=3.35in \epsfbox{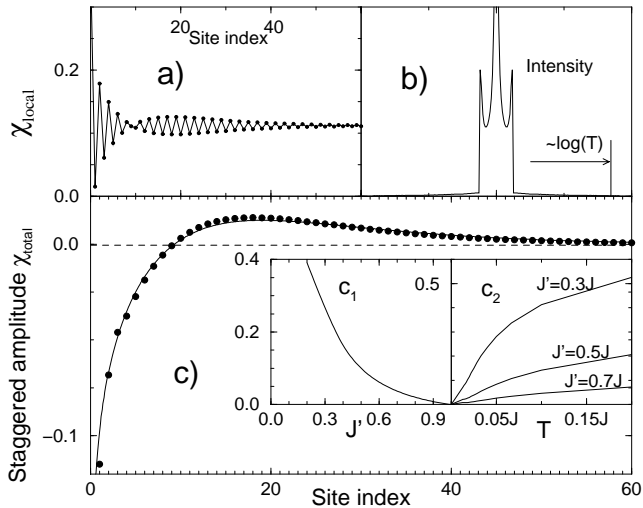}}
\end{center}
\caption{a) The local susceptibilities as a function of site index for $T=0.05J$
and $J' = 0.7J$. b) The corresponding
typical NMR spectrum. 
c) The fit of the alternating amplitude to Eq.~(\ref{chi})
with the appropriate coefficients (inset). 
\label{knightshift}}
\end{figure}

For the impurity in question we can predict a similar effect which then 
could again be measured by NMR experiments.  The leading operator
of the impurity spin also induces a staggered component which we have
explicitly calculated to be
\begin{equation} 
\chi(x) \ \propto \ (-1)^x \log{\left[
\tanh{ (x T)}\right]}.
\label{chiper}
\end{equation}
For small distances $x$ this formula shows the logarithmic temperature
dependence of the impurity susceptibility explicitly.
Clearly, for small coupling $J'$ and at larger temperatures we expect 
the open chain behavior, but as the temperature
is lowered (or the coupling is increased) the alternating part in
Eq.~(\ref{chiper}) will start to dominate, which has the opposite
sign.  This competition between the two contribution is indeed
observed in Fig.~\ref{knightshift}. Moreover,  below $T_K$ the total staggered
contribution always fits extremely well to a superposition
\begin{equation}
\chi_{\rm total}(x) \ = \ c_1 \log[\tanh(x T)] 
\  + \ c_2 \frac{x \sqrt{T}}{\sqrt{\sinh 4 x T}}. \label{chi}
\end{equation}

We have determined the coefficients $c_1$ and $c_2$ for all coupling
strengths $J' \ge 0.2J$ in Fig.~\ref{knightshift}, where $c_1$ is
temperature independent and $c_2$ renormalizes to zero if the temperature
is lowered as expected.  We have also calculated the corresponding NMR
spectrum which shows a logarithmically broad feature and distinctive kinks,
which have a unique temperature dependence
as determined from Eq.~(\ref{chi}) depending on the coupling strength $J'$.
These quantitative predictions should allow experiments
to observe two-channel Kondo physics by NMR and susceptibility measurements
on doped spin-1/2 chain compounds.

\end{document}